\documentclass[twocolumn,showpacs,preprintnumbers,amssymb,aps,pra]{revtex4}
\usepackage{graphicx}
\usepackage{dcolumn}
\usepackage{bm}
\usepackage{latexsym,epsfig}

\usepackage{array}
\usepackage{amstext}

\begin{document}

\title{Theoretical studies of the long lifetimes of the $6d \ ^2D_{3/2,5/2}$ states in Fr: Implications for parity nonconservation measurements}
\vspace*{0.5cm}

\author{B. K. Sahoo \footnote{Email: bijaya@prl.res.in}}
\affiliation{Theoretical Physics Division, Physical Research Laboratory, Ahmedabad-380009, India}
\author{B. P. Das}
\affiliation{International Education and Research Center of Science and Department of Physics, Tokyo Institute of Technology, 2-1-2-1-H86 Ookayama
Meguro-ku, Tokyo 152-8550, Japan}
   
\date{Recieved date; Accepted date}
\vskip1.0cm

\begin{abstract}
\noindent
The lifetimes of the $6d \ ^2D_{3/2}$ and $6d \ ^2D_{5/2}$ states in Fr are estimated to be 540(10) ns and 1704(32) ns respectively. They are 
determined by calculating the radiative transition amplitudes of the allowed electric dipole (E1) and the forbidden electric quadrupole (E2) 
and magnetic dipole (M1) channels using the second order many-body perturbation theory (MBPT(2)) and the coupled-cluster (CC) method at different 
levels of approximation in the relativistic framework. These long lifetimes and the large electric dipole parity non conserving amplitudes of 
$7s \ ^2S_{1/2} \rightarrow 6d \ ^2D_{3/2,5/2}$ transitions strongly favour Fr as a leading candidate for the measurement of parity 
nonconservation arising from the neutral current weak interaction and the nuclear anapole moment.
\end{abstract}

\maketitle

Francium (Fr) is considered to be a promising candidate for the measurements of the electric dipole moment (EDM) arising due to the violations 
of parity and time reversal symmetries \cite{sakemi, stancari}, parity nonconservation (PNC) effects due to the neutral weak interaction 
\cite{stancari,dreischuh} and the nuclear anapole moment \cite{gomez0,gomez1} as it is the heaviest alkali metal atom. The focus of all the 
ongoing Fr PNC experiments is the $7s \ ^2S_{1/2} \rightarrow 8s \ ^2S_{1/2}$ transition \cite{stancari,dreischuh,gomez1}; which is largely
inspired by the Cs PNC experiment \cite{wood}. However, relativistic many-body calculations show that the  PNC amplitudes in the $7s \ ^2S_{1/2} 
\rightarrow 6d \ ^2D_{3/2,5/2}$ transition amplitudes in Fr are about three times larger than that of the $7s \ ^2S_{1/2} \rightarrow 
8s \ ^2S_{1/2}$ transition \cite{bijaya4,roberts}. The $S-D$ PNC transitions in the singly ionized Ba, Ra and Yb have been the subject of 
theoretical investigations \cite{bijaya1,wansbeek,bijaya2} and the principle of their measurements has been discussed \cite{fortson,mandal}. It 
has also been highlighted that the PNC measurements for the $S-D_{5/2}$ transitions in these ions would provide unambiguous signatures of the 
existence of the nuclear anapole moment (NAM) \cite{geetha,bijaya3}, which is still an open question \cite{haxton}. Apart from exhibiting large PNC 
effects, another important aspect of these transitions is that the  excited $D$ states  in theses ions are  metastable stables, and they provide 
long interrogation times which enhances the precision of the measurement of the small PNC effects \cite{fortson}. In this Rapid Communication, 
we present the results of our theoretical studies on the lifetimes the $6d \ ^2D_{3/2}$ and $6d \ ^2D_{5/2}$ states in Fr, which were undertaken
to assess the feasibility of the measurement of PNC in this atom using the  $7s \ ^2S_{1/2} \rightarrow 6d \ ^2D_{3/2,5/2}$ transitions.

 An electron from the $6d \ ^2D_{3/2}$ state can decay to its low-lying $7p \ ^2P_{1/2}$ and  $7p \ ^2P_{3/2}$ states by the electric dipole (E1) 
and forbidden magnetic octupole (M3) transitions and to the ground state by the forbidden magnetic dipole (M1) and electric quadrupole (E2)
transitions. We neglect contributions due to the M3 transition as the corresponding transition probability is very weak owing to its inversely 
proportional to seventh power of transition wavelength. Similarly, an electron from the $6d \ ^2D_{5/2}$ state can decay to its fine-structure 
partner $6d \ ^2D_{3/2}$ state via both the M1 and E2 transitions while to the low-lying $7p \ ^2P_{3/2}$ state by the E1 transition and to the ground 
state by the E2 transition. In this case too, we have omitted contributions due to the M3 transition. The transition probabilities due to the 
above E1, E2 and M1 channels for a transition, say, $\vert \Psi_i \rangle \rightarrow \vert \Psi_f \rangle$ are given by 
\begin{eqnarray}
&& A^{E1}_{if} = \frac{2.0261\times 10^{-6}}{\lambda_{if}^3 g_i} S_{if}^{E1} \label{eqn5} \\
&& A^{E2}_{if} = \frac{1.1195\times 10^{-22}}{\lambda_{if}^5 g_i} S_{if}^{E2} \label{eqn6} \\
\text{and}  && \nonumber \\
&& A^{M1}_{if} =  \frac{2.6971\times 10^{-11}}{\lambda_{if}^3 g_i} S_{if}^{M1}, \label{eqn7} \ \ \ \ \ \ \
\end{eqnarray}
where the quantity $S^O_{if} = \mid {\langle \Psi_i \vert \vert O \vert \vert \Psi_f \rangle} \mid^2$ is known as the line strength for the 
corresponding reduced matrix element $\mid {\langle \Psi_i \vert \vert O \vert \vert \Psi_f \rangle} \mid$ of a transition operator $O$. These 
quantities are later given in atomic unit (a.u.). In the above expressions, $g_i=2J_i+1$ is the degeneracy factor of the state $\vert \Psi_i 
\rangle$ with the angular momentum $J_i$ and the transition wavelength ($\lambda_{if}$) is used in nm which when substituted, the transition 
probabilities ($A^O_{if}$s) are obtained in $s^{-1}$. The lifetime ($\tau$) of the atomic state $\vert \Psi_i \rangle$ is determined by taking 
the reciprocal of the total emission transition probabilities involving all the possible spontaneous transition channels (in $s$). i.e. 
\begin{eqnarray}
\tau_i &=& \frac {1} {\sum_{O,f} A^{O}_{if}},
\label{eqn8}
\end{eqnarray}
where the summations over $O$ and $f$ correspond to all probable decay channels and all the lower states respectively. We have attempted to
obtain accurate results for the transition probabilities, hence the lifetimes of the atomic states by performing relativistic many-body calculations
of the line strengths  and using wavelengths that are determined from the experimental transition energies given
in the National Institute of Science and Technology (NIST) database \cite{nist}. 

To investigate the role of the electron correlation effects in the evaluation of the radiative transition amplitudes, we employ the second order many-body 
perturbation theory (MBPT(2)) and the coupled-cluster (CC) method in the relativistic framework. Further, we take different levels of 
approximation in the CC method to see the convergence in the results. We give below a brief description of these methods using 
Bloch's prescription \cite{lindgren}, in which atomic wave function of state $\vert \Psi_n \rangle$ is expressed as 
\begin{eqnarray}
 \vert \Psi_n \rangle = \Omega_n \vert \Phi_n \rangle,
\end{eqnarray}
where $\Omega_n$ and $\vert \Phi_n \rangle$ are known as the wave operator and reference state respectively. The ground and the 
considered excited $6D$ states of Fr have the electronic configurations as $[6p^6] 7s \ ^2S_{1/2}$ and $[6p^6] 6d \ ^2D_{3/2,5/2}$ respectively.
To reduce the computational effort, we construct these states by creating a common reference state function $\vert \Phi_c \rangle$ 
with the $[6p^6]$ configuration using the Dirac-Hartree-Fock (DHF) method. In this approach the atomic Hamiltonian ($H$) in the 
Dirac-Coulomb interaction approximation is divided as DHF Hamiltonian $H_0$ and residual Coulomb interaction $V_r$. For the calculation of the 
exact states with a valence orbital, we define new working reference states as $\vert \Phi_n \rangle = a_n^{\dagger} \vert \Phi_c 
\rangle$. Here $a_n^{\dagger}$ appends an electron from the respective valence orbital denoted by an index $n$. As a consequence $\Omega_n$ 
can now be divided as
\begin{eqnarray}
 \Omega_n =  1+ \chi_c + \chi_n ,
\end{eqnarray}
where $\chi_c$ and $\chi_n$ are responsible for carrying out excitations (generating configuration state functions) from 
$\vert \Phi_c \rangle$ and $\vert \Phi_n \rangle$, respectively, due to $V_r$. In a perturbative series expansion, we have
\begin{eqnarray}
 \chi_c = \sum_k \chi_c^{(k)} \ \ \text{and} \ \ \chi_n=\sum_k \chi_n^{(k)}.
\end{eqnarray}
In these expressions, the superscripts imply number of $V_r$ considered in the calculations and represents order of perturbation; e.g. 
MBPT(2) method bears terms up to two $V_r$ ($k=2$). 

Using the generalized Bloch's equation, $k$th order amplitudes for the $\chi_c$ and $\chi_n$ operators are obtained by \cite{lindgren}
\begin{eqnarray}
 [\chi_c^{(k)},H_0]P &=& Q V_r(1+ \chi_c^{(k-1)} )P 
\end{eqnarray}
and
\begin{eqnarray}
[\chi_n^{(k)},H_0]P &=& QV_r (1+ \chi_c^{(k-1)}+ \chi_n^{(k-1)}) P 
 - \sum_{m=1 }^{k-1}\chi_n^{(k-m)} \nonumber \\ && \times PV_r(1+\chi_c^{(m-1)}+\chi_n^{(m-1)})P, 
\end{eqnarray}
where the projection operators $P=\vert \Phi_c \rangle \langle \Phi_c \vert $ and $Q= 1- P$ describe the model space and the orthogonal space 
of the Hamiltonian $H_0$ respectively. Note that here $\chi_c^{(0)}=0$ and $\chi_n^{(0)}=0$. Using these amplitudes, the energy 
of the state $\vert \Psi_n \rangle$ is evaluated by using an effective Hamiltonian
\begin{eqnarray}
 H_n^{eff}= P H\Omega_n P. 
\end{eqnarray}

\begin{table*}[t]
\caption{Magnitudes of the reduced matrix elements $\langle J_i ||O|| J_f  \rangle$ of transition operators ($O$s) given in a.u. from different 
many-body methods. Uncertainties from the finite size basis set, non-inclusion of the Breit interaction and due to QED effects are quoted 
using the MBPT(2) method. Wavelengths ($\lambda_{if}$) from the NIST database \cite{nist} are quoted in nm for the respective transitions.}
\begin{ruledtabular}
\begin{tabular}{lccccccccccc} 
  $J_i \rightarrow J_f$                       & $\lambda_{if}$ & DHF    & MBPT(2) & LCCSD & CCSD$^{(2)}$ & CCSD$^{(4)}$  & CCSD$^{(\infty)}$ & CCSD(T) & \multicolumn{3}{c}{Uncertainties} \\  
 \cline{10-12} \\ 
  &                &       &         &       &              &               &                   &         & Basis & Breit & QED         \\
 \hline 
 & & \\
 \multicolumn{12}{c}{E1 matrix elements} \\  
 $6d \ ^2D_{3/2} \rightarrow 7p \ ^2P_{1/2}$  & 2504.7    &  9.22 &  7.73   &  6.81 &  7.46   &  7.47  &  7.47  &  7.43  & 0.05  & $-0.01$ &  0.001\\
 $6d \ ^2D_{3/2} \rightarrow 7p \ ^2P_{3/2}$  & 4336.7    &  4.28 &  3.57   &  3.12 &  3.44   &  3.45  &  3.45  &  3.42  & 0.02  & $-0.01$ & $\sim 0.0$ \\
 $6d \ ^2D_{5/2} \rightarrow 7p \ ^2P_{3/2}$  & 3991.0    & 12.80 & 10.83   &  9.68 & 10.54   & 10.55  & 10.55  & 10.51  & 0.07  & $-0.02$ & 0.001 \\
 & & \\
 \multicolumn{12}{c}{E2 matrix elements} \\  
 $6d \ ^2D_{3/2} \rightarrow 7s \ ^2S_{1/2}$  & 616.5     & 43.10 & 33.74   & 31.39 &  34.02  & 34.06  & 34.06 & 33.78  &  0.20  & $-0.03$ & $-0.02$ \\  
 $6d \ ^2D_{5/2} \rightarrow 7s \ ^2S_{1/2}$  & 608.7     & 52.74 & 41.69   & 39.31 &  42.24  & 42.27  & 42.27 & 41.96  &  0.24  & $-0.04$ & $-0.02$ \\
 $6d \ ^2D_{5/2} \rightarrow 6d \ ^2d_{3/2}$  & 50057.6   & 47.70 & 32.14   & 27.18 &  32.01  & 32.03  & 32.03 & 31.49  &  0.55  & $-0.16$ &  0.01  \\
 & & \\
 \multicolumn{12}{c}{M1 matrix elements} \\ 
 $6d \ ^2D_{3/2} \rightarrow 7s \ ^2S_{1/2}$  & 616.5     & $\sim 0.0$ & 0.001 & 0.002 & 0.001 & 0.001 & 0.001 & 0.001  & $\sim 0.0$  &  $\sim 0.0$ & $\sim 0.0$ \\  
 $6d \ ^2D_{5/2} \rightarrow 6d \ ^2d_{3/2}$  & 50057.6   & 1.549      & 1.550 & 1.548 & 1.547 & 1.552 & 1.552 & 1.552  & $\sim 0.0$  &  $0.001$ & $\sim 0.0$  \\
\end{tabular}
\end{ruledtabular}
\label{tab1}
\end{table*}
Using the CC ansatz, the above expressions can be put together to construct a wave operator to infinite order as 
\begin{eqnarray}
 \vert \Psi_n \rangle &=& \Omega_n \vert \Phi_n = e^T \{ 1+ S_n \} \vert \Phi_n \rangle,
 \label{eqcc}
\end{eqnarray}
such that $\chi_c= e^T-1$ and $\chi_n=e^TS_n -1$. Here the $T$ and $S_n$ are the CC excitation operators that excite electrons from the core 
and core along with the valence orbital respectively. In this work, we have only accounted the singly and doubly excited states denoting the CC 
operators by subscripts $1$ and $2$ respectively as
\begin{eqnarray}
 T=T_1 +T_2 \ \ \ \text{and} \ \ \ S_n = S_{1n} + S_{2n}.
\end{eqnarray}
This is referred to as the CCSD method in the literature. When only the linear terms are retain in Eq. (\ref{eqcc}) with singles and doubles 
approximation, it is referred to as the LCCSD method. The amplitudes of the $T$ and $S_n$ operators are determined using the expressions
\begin{eqnarray}
 H_N \chi_c P = Q H_N P 
\end{eqnarray}
and
\begin{eqnarray}
H_N \chi_n P &=& QH_N (1+\chi_c) P - \chi_n H_N^{eff}, 
\end{eqnarray}
where we have defined normal order Hamiltonian $H_N=H-PHP$ and the effective Hamiltonian to evaluate ionization potential (IP) of an electron 
from the valence orbital $n$ of the respective state is given by $H_N^{eff}=PH_N(1+\chi_c+\chi_n)P$. We have also included contributions from 
important triples excitations perturbatively from $\chi_c^{(2)}$ and $\chi_n^{(2)}$ in the construction of $H_N^{eff}$ and the approach is 
referred to as the CCSD(T) method.

  After obtaining amplitudes using the above equations, the transition matrix element of an operator $O$ between the states
$\vert \Psi_i \rangle$ and $\vert \Psi_f \rangle$ is evaluated using the expression
\begin{eqnarray}
\frac{\langle \Psi_f \vert O \vert \Psi_i \rangle}{\sqrt{\langle \Psi_f \vert \Psi_f\rangle \langle \Psi_i \vert \Psi_i\rangle}} 
&=& \frac {\langle \Phi_f \vert \Omega_f^{\dagger} O \Omega_i \vert \Phi_i\rangle}
{\sqrt{ \langle \Phi_f \vert \Omega_f^{\dagger} \Omega_f \vert \Phi_f\rangle \langle \Phi_i \vert \Omega_i^{\dagger} \Omega_i \vert \Phi_i\rangle} }  . \ \ \ \ \ 
\label{preq}
\end{eqnarray}
This gives rise to a finite number of terms for the MBPT(2) and LCCSD methods, but it involves two non-terminating series in the numerator and 
denominator which are $e^{T^{\dagger}} O e^T$ and $e^{T^{\dagger}} e^T$ respectively in the CCSD and CCSD(T) methods. In order to evaluate all
the significant contributions from these series, we have used the Wick's generalized theorem \cite{lindgren} to divide these terms into the 
effective one-body, two-body and three-body terms. The effective one-body terms are the dominant ones, they are computed first considering the CC 
terms with the approximations $e^{T^{\dagger}} O e^T \simeq O+OT + T^{\dagger}O + \frac{1}{2}OT^2 +  \frac{1}{2}T^{\dagger 2}O+ T^{\dagger}OT$ and 
$e^{T^{\dagger}} e^T \simeq T^{\dagger}T + \frac{1}{2}T^{\dagger} T^2 + \frac{1}{2}T^{\dagger 2} T$. Then, they are stored and contracted with 
the $T_2$ and $T_2^{\dagger}$ operators avoiding repetitions of the diagrams in a self-consistent procedure to account for the higher order 
one-body terms from the non-terminating series. They are again stored as an intermediate form for the further contraction with the $S_n$ and 
$S_n^{\dagger}$ operators. Similarly the effective two-body and three-body terms are computed after contracting with the above effective one-body 
terms with the $T_2$ and $T_2^{\dagger}$ operators, but they are computed directly contracting with the $S_n$ and $S_n^{\dagger}$ operators. 
Thus, these effective two-body and three-body terms also have contributions from the non-terminating series. To see the convergence of the 
results with the series expansion, we present contributions with $k$ numbers of $T$ and/or $T^{\dagger}$ operators from these non-terminating 
series, which we refer to as the CCSD$^{(k)}$ method considering terms up to $k \rightarrow \infty$ in a self-consistent procedure as described 
above. Our final CCSD results correspond to the CCSD$^{(\infty)}$ method. The same procedure is also adopted for the CCSD(T) method. The 
contribution from the normalizations of the wave functions (${\cal C}_{norm}$) is estimated explicitly using the expression
\begin{eqnarray}
 {\cal C}_{norm} = \left [ \frac{\langle \Psi_f \vert O \vert \Psi_i \rangle}{\sqrt{\langle \Psi_f \vert \Psi_f\rangle \langle \Psi_i \vert \Psi_i\rangle}} 
 - \langle \Psi_f \vert O \vert \Psi_i \rangle \right ].
\end{eqnarray}

\begin{table*}[t]
\caption{Contributions to the reduced matrix elements $\langle J_i ||O|| J_f  \rangle$ from various terms of the CCSD method (in a.u.). 
Differences between these values from the CCSD results are quoted in Table \ref{tab1} correspond to those non-linear terms that are not mentioned 
explicitly here.}
\begin{ruledtabular}
\begin{tabular}{lccccccccccc} 
  $J_i \rightarrow J_f$  & $O$ & $OT_1$ & $T_1^{\dagger}O$ & $OS_{1f}$ & $S_{1i}^{\dagger}O$  & $S_{1i}^{\dagger}OS_{1f}$  &  $OS_{2f}$ & $S_{2i}^{\dagger}O$ &  $S_{1i}^{\dagger}OS_{2f}$ & $S_{2i}^{\dagger}OS_{1f}$ & ${\cal C}_{norm}$ \\  
 \hline 
 & & \\
 \multicolumn{12}{c}{E1 matrix elements} \\  
 $6d \ ^2D_{3/2} \rightarrow 7p \ ^2P_{1/2}$  &  9.22 & $\sim 0.0$ &  0.019 & $-0.437$  & $-0.877$  & 0.246  & $-0.208$  & $-0.248$  & $-0.020$ & $-0.020$ & $-0.324$ \\
 $6d \ ^2D_{3/2} \rightarrow 7p \ ^2P_{3/2}$  &  4.28 & $\sim 0.0$ &  0.002 & $-0.132$  & $-0.522$  & 0.089  & $-0.081$  & $-0.097$  & $-0.005$ & $-0.007$ & $-0.139$ \\
 $6d \ ^2D_{5/2} \rightarrow 7p \ ^2P_{3/2}$  & 12.80 & $\sim 0.0$ &  0.006 & $-0.381$  & $-1.372$  & 0.238  & $-0.233$  & $-0.292$  & $-0.008$ & $-0.020$ & $-0.342$ \\
 & & \\
 \multicolumn{12}{c}{E2 matrix elements} \\  
 $6d \ ^2D_{3/2} \rightarrow 7s \ ^2S_{1/2}$  & 43.10 & $\sim 0.0$ & 0.013 & $-6.306$  & $-2.683$  & 1.547 & $-0.139$  &  $-0.111$ & $-0.027$ & $-0.186$ & $-1.618$ \\  
 $6d \ ^2D_{5/2} \rightarrow 7s \ ^2S_{1/2}$  & 52.74 & $\sim 0.0$ & 0.015 & $-7.644$  & $-2.871$  & 1.666 & $-0.125$  &  $-0.148$ & $-0.017$ &  0.004 & $-1.6$ \\
 $6d \ ^2D_{5/2} \rightarrow 6d \ ^2d_{3/2}$  & 47.70 & $\sim 0.0$ & $\sim 0.0$ & $-8.994$  & $-8.116$  & 3.264 & $-0.079$  & $-0.010$ & $-0.012$ & $-0.021$ & $-1.947$ \\
 & & \\
 \multicolumn{12}{c}{M1 matrix elements} \\ 
 $6d \ ^2D_{3/2} \rightarrow 7s \ ^2S_{1/2}$  & $\sim 0.0$ & $\sim 0.0$ & $\sim 0.0$ & $\sim 0.0$ & $\sim 0.0$ & 0.0004 & 0.0005  & 0.0001 &  $\sim 0.0$ & $\sim 0.0$ & $\sim 0.0$ \\  
 $6d \ ^2D_{5/2} \rightarrow 6d \ ^2d_{3/2}$  & 1.549  & $\sim 0.0$ & $\sim 0.0$ & 0.003 & $-0.002$ & 0.078 & 0.003  & $\sim 0.0$  &  $\sim 0.0$ & $\sim 0.0$ & $-0.093$ \\
\end{tabular}
\end{ruledtabular}
\label{tab2}
\end{table*}

In Table \ref{tab1}, we give the radiative transition matrix elements for all the considered channels from the DHF, MBPT(2), LCCSD, CCSD and 
CCSD(T) methods to analyze the propagation of the correlation effects through various levels of approximations in the many-body theories 
and the experimental values of the transition wavelengths from the NIST database \cite{nist} that we have used later. We also 
give contributions from the CCSD method by truncating the non-linear terms with $k=2$, $k=4$ and from a self-consistent ($k=\infty$) calculation. For $k=2$, 
the expression evaluating the property given by Eq. (\ref{preq}) has the same number of terms as does the LCCSD method. Therefore, differences in the 
results from the LCCSD and CCSD$^{(2)}$ methods imply the correlation contributions arising through the non-linear terms in the wave function 
determining equations of the CCSD method and are found to be quite large. Often, these contributions are neglected in the calculations as they require prohibitively
large computational resources for their evaluation. Again, we observe from the trends that the correlation effects at the MBPT(2) method  are large, and that there are strong cancellations in the LCCSD approximation and the results almost converge for $k=4$ when
non-linear terms are included in the CCSD method. The discrepancy in the results of the CCSD$^{(4)}$ and CCSD$^{(\infty)}$ methods is beyond
second significant digit implying that the results have converged within the precision of our interest. The valence triple excitations seem to change the 
results slightly. We also give uncertainties associated with these results by estimating contributions due to the finite size of our basis set, 
neglected contributions from the Breit interaction and corrections from the quantum electrodynamics (QED). These estimates are carried out 
using the MBPT(2) method which gives the largest correlation effects. 

\begin{table}[t]
\caption{Transition probabilities ($A_{if}^O$) due to different transition decay channels ($O$s) in $s^{-1}$ and their branching ratios 
from the $6d \ ^2D_{3/2}$ and $6d \ ^2D_{5/2}$ states of Fr. Uncertainties are quoted within the parentheses.}
\begin{ruledtabular}
\begin{tabular}{lccc} 
  $J_i \rightarrow J_f$                       & $O$ & $A_{if}^O $ & Branching ratio  \\  
 \hline 
 & & \\ 
 $6d \ ^2D_{3/2} \rightarrow 7p \ ^2P_{1/2}$  & E1 & 1779511(33688) & 0.96 \\
 $6d \ ^2D_{3/2} \rightarrow 7p \ ^2P_{3/2}$  & E1 & 72637(1280) & 0.04 \\
 $6d \ ^2D_{3/2} \rightarrow 7s \ ^2S_{1/2}$  & E2 & 35.96(60) & $\sim 0.0$ \\
 $6d \ ^2D_{3/2} \rightarrow 7s \ ^2S_{1/2}$  & M1 & $\sim 0.0$ & $\sim 0.0$ \\
 & & \\
 $6d \ ^2D_{5/2} \rightarrow 7p \ ^2P_{3/2}$  & E1 & 586776(11219) & $\sim 1.0$ \\ 
 $6d \ ^2D_{5/2} \rightarrow 7s \ ^2S_{1/2}$  & E2 & 39.33(19) & $\sim 0.0$ \\ 
 $6d \ ^2D_{5/2} \rightarrow 6d \ ^2d_{3/2}$  & E2 & $\sim 0.0$ & $\sim 0.0$ \\
 $6d \ ^2D_{5/2} \rightarrow 6d \ ^2d_{3/2}$  & M1 & $\sim 0.0$ & $\sim 0.0$ \\
\end{tabular}
\end{ruledtabular}
\label{tab3}
\end{table}

 After analyzing the trends in the correlation effects at different levels of approximation, we now focus on the contributions from different 
terms in the CCSD method. We present these results in Table \ref{tab2} along with the contributions from ${\cal C}_{norm}$. Contributions from 
the corresponding radiative operator $O$ are the DHF results, $OT_{1}$ and its complex conjugate terms give core-valence correlations, 
$OS_{1f}$ and $S_{1i}^{\dagger}O$ give the pair correlation effects involving the valence orbitals, $OS_{2f}$ and $S_{2i}^{\dagger}O$ give the 
core-polarization correlation effects involving the valence orbitals, etc. Contributions from the other non-linear terms such as representing the 
core pair correlation effects coming through the $T_2^{\dagger}OT_2$ term are not given explicitly in the above table, however their 
contributions can be obtained by taking the differences of the contributions given in Table \ref{tab2} and the final CCSD results given in 
Table \ref{tab1}. As can be seen from Table \ref{tab2}, core-valence correlations are small and the largest correlation effects come through 
the pair correlation effects in the E1 and E2 matrix element calculations. Nevertheless, the core-polarization effects are also very significant
and are the dominant ones in the calculations of the M1 matrix elements. We also find contributions from ${\cal C}_{norm}$ to be fairly large.

We now use the matrix elements from the CCSD(T) method and the experimental wavelengths mentioned in Table \ref{tab1} to evaluate the transition 
probabilities due to different radiative decay channels from the $6d \ ^2D_{3/2}$ and $6d \ ^2D_{5/2}$ states. These values are given in Table 
\ref{tab3} along with their branching ratios for the individual transitions. As can be seen from this table that branching ratios are dominated 
by the E1 transitions and they are entirely responsible for determining the lifetimes of the $6d \ ^2D_{3/2}$ and $6d \ ^2D_{5/2}$ states. Using 
these transition probabilities, we estimate the lifetimes of the $6d \ ^2D_{3/2}$ and $6d \ ^2D_{5/2}$ states as $\tau_{6d3/2}=540(10)$ ns and 
$\tau_{6d5/2}=1704(32)$ ns, respectively.  These values are very large compared to the other low-lying excited $7p \ ^2P_{1/2,3/2}$, 
$8s \ ^2S_{1/2}$, $7p \ ^2P_{1/2,3/2}$, $7d \ ^2D_{3/2,5/2}$ and $9s \ ^2S_{1/2}$ states of Fr which are measured till date as $\tau_{7p1/2}=29.45(11)$ ns,
$\tau_{7p3/2}=21.02(15)$ ns, $\tau_{8s1/2}=53.30(44)$ ns, $\tau_{7d3/2}=73.6(3)$ ns and $\tau_{7d5/2}=67.7(2.9)$ ns, $\tau_{8p1/2}=149.3(3.5)$ ns, 
$\tau_{8p3/2}=83.5(1.5)$ ns and $\tau_{9s1/2}=107.53(90)$ ns respectively \cite{gomez2}. Given the large PNC amplitudes in the $7s \ ^2S_{1/2} 
\rightarrow 6d \ ^2D_{3/2,5/2}$ transitions \cite{roberts} compared to the $7s \ ^2S_{1/2} \rightarrow 8s \ ^2S_{1/2}$ transition in Fr 
\cite{bijaya4} and with the corresponding transitions in Ra$^+$ ion \cite{wansbeek,bijaya3} and the long lifetimes of the excited $6d \ ^2D_{3/2}$ 
and $6d \ ^2D_{5/2}$ states, make Fr a potentially attractive candidate for a PNC experiment.

 In summary, we have performed relativistic many-body calculations of the lifetimes of the $6d \ ^2D_{3/2}$ and $6d \ ^2D_{5/2}$ states of Fr and 
 they are found to be  
large. These results  favour the measurement of PNC in the $7s \ ^2S_{1/2} \rightarrow 6d \ ^2D_{3/2}$ and $7s \ ^2S_{1/2} 
\rightarrow 6d \ ^2D_{5/2}$ transitions of Fr, where calculations predict large PNC effects. For the evaluation of the lifetimes, we have calculated radiative transition matrix elements using the relativistic
CC method. We have also investigated the roles of the electron correlation effects in the determination of these quantities systematically by 
approximating many-body methods at different levels and give contributions explicitly from various CCSD terms.
 
This work was supported partly by INSA-JSPS under the project no. IA/INSA-JSPS Project/2013-2016/February 28,2013/4098. 
Computations were carried out using the Vikram-100TF HPC cluster at the Physical Research Laboratory, Ahmedabad, India.


\begin{thebibliography}{22}
\bibitem{sakemi}
Y. Sakemi, K.Harada, T. Hayamizu, M. Itoh, H. Kawamura, S. Liu, H. S. Nataraj, A. Oikawa, M. Saito, T. Sato, 
H. P. Yoshida, T. Aoki, A. Hatakeyama, T. Murakami, K. Imai, K. Hatanaka, T. Wakasa, Y. Shimizu, and M. Uchida, 
J. Phys.: Conf. Series {\bf 302}, 012051 (2011).
\bibitem{stancari}
G. Stancari, S. N. Atutov, R. Calabrese, L. Corradi, A. Dainelli, C. de Mauro, A. Khanbekyan, E. Mariotti, 
P. Minguzzi, L. Moi, S. Sanguinetti, L. Tomassetti and S. Veronesi, Eur. Phys. J. Special Topics {\bf 150}, 389 (2007).
\bibitem{dreischuh}
T. Dreischuh, E. Taskova, E. Borisova and A. Serafetinides, {\it 15th International School on Quantum Electronics: Laser Physics and
Applications}, Proceedings SPIE, vol. {\bf 7027}, p. 70270C-70270C-7 (2008).
\bibitem{gomez0}
E. Gomez, S. Aubin, R. Collister, J. A. Behr, G. Gwinner, L. A. Orozco, M. R. Pearson, M. Tandecki, D. Sheng and J. Zhang, J. Phys.: Conf. Series {\bf 387} 012004 (2012).
\bibitem{gomez1}
E. Gomez, S. Aubin, G. D. Sprouse, L. A. Orozco and D. P. DeMille, Phys. Rev. A {\bf 75}, 033418 (2007).
\bibitem{wood}
C. S. Wood, S. C. Bennett, D. Cho, B. P. Masterson, J. L. Roberts, C. E. Tanner and C. E. Wieman, Science {\bf 275}, 1759 (1997).
\bibitem{bijaya4}
B. K. Sahoo, J. Phys. B: At. Mol. Opt. Phys. {\bf 43}, 085005 (2010).
\bibitem{roberts}
B. M. Roberts, V. A. Dzuba, and V. V. Flambaum, Phys. Rev. A {\bf 89}, 012502 (2014).
\bibitem{bijaya1}
B. K. Sahoo, R. Chaudhuri, B. P. Das and D. Mukherjee, Phys. Rev. Lett. {\bf 96}, 163003 (2006).
\bibitem{wansbeek}
L. W. Wansbeek, B. K. Sahoo, R. G. E. Timmermans, K. Jungmann, B. P. Das and D. Mukherjee, Phys. Rev. A {\bf 78}, 050501(R) (2008).
\bibitem{bijaya2}
B. K. Sahoo and B. P. Das, Phys. Rev. A {\bf 84}, 010502(R) (2011).
\bibitem{fortson}
N. Fortson, Phys. Rev. Lett. {\bf 70}, 2383 (1993).
\bibitem{mandal}
P. Mandal and M. Mukherjee, Phys. Rev. A {\bf 82}, 050101(R) (2010).
\bibitem{geetha}
K. P. Geetha, Angom Dilip Singh, B. P. Das and C. S. Unnikrishnan, Phys. Rev. A {\bf 58}, R16(R) (1998).
\bibitem{bijaya3}
B. K. Sahoo, P. Mandal and M. Mukherjee, Phys. Rev. A {\bf 83}, 030502(R) (2011).
\bibitem{haxton}
W. C. Haxton and C. E. Wieman, Annu. Rev. Nucl. Part. Sci. {\bf 51}, 261 (2001).
\bibitem{fedosseev}
V. N. Fedosseev, Y. Kudryavtsev and V. I. Mishin, Phys. Scr. {\bf 85}, 058104 (2012).
\bibitem{koester}
U. K\"oster, V. N. Fedoseyev and  V. I. Mishin, Spectrochimica Acta Part B: Atomic Spectroscopy {\bf 58}, 967 (2003).
\bibitem{nist}
http://physics.nist.gov/PhysRefData/ASD/levels\_form.html
\bibitem{lindgren} 
I. Lindgren and J. Morrison, {\it Atomic Many-Body Theory}, Second Edition, Springer-Verlag, Berlin, Germany (1986).
\bibitem{gomez2}
E. Gomez, L. A. Orozco and G. D. Sprouse, Rep. Prog. Phys. {\bf 69}, 79 (2006).

\end{thebibliography}
\end{document}